\documentclass[aps,prl,twocolumn,superscriptaddress]{revtex4}
\usepackage{graphicx}

\begin{document}
\include{srctex}
\bibliographystyle{apsrev}
\title{A New Nucleosynthesis Constraint on the Variation of G}
\author{Craig J. Copi}
\affiliation{Department of Physics, Case Western Reserve University,
  Cleveland, OH 44106-7079}
\author{Adam N. Davis}
\affiliation{Department of Physics, Case Western Reserve University,
  Cleveland, OH 44106-7079}

\author{Lawrence M. Krauss}
\affiliation{Department of Physics, Case Western Reserve University,
  Cleveland, OH 44106-7079}
\affiliation{Department of Astronomy, Case Western Reserve University,
  Cleveland, OH 44106-7079}
\date{\today}

\begin{abstract}
  Big Bang Nucleosynthesis can provide, via constraints on the expansion
  rate at that time, limits on possible variations in Newton's Constant,
  $G$.  The original analyses were performed before an independent
  measurement of the baryon-to-photon ratio from the cosmic microwave
  background was available.  Combining this with recent measurements of the
  primordial deuterium abundance in quasar absorption systems now allows
  one to derive a new tighter constraint on $G$ without recourse to
  considerations of helium or lithium abundances.  We find that, compared
  to todays value, $G_0$, $G_{BBN}/G_0=1.01^{+0.20}_{-0.16}$ at the
  $68.3$\% confidence level.  If we assume a monotonic power law time
  dependence, $G\propto t^{-\alpha}$, then the constraint on the index is
  $-0.004 < \alpha < 0.005$. This would translate into
  $-3\times10^{-13}\;\textrm{yr}^{-1} < \left(\dot
    G/G\right)_{\textrm{today}} < 4 \times 10^{-13}\;\textrm{yr}^{-1}$.
\end{abstract}
\maketitle

The predictions of the light element abundances from big bang
nucleosynthesis (BBN) has long served as a powerful probe of the early
Universe. These predictions depend on a number of parameters such as the
number of light particle species in equilibrium at that time, the baryon to
photon ratio ($\eta$), nuclear reaction rates, and the gravitational
constant, $G$. BBN constraints on cosmological parameters have been
traditionally limited by uncertainties in the observed primordial
abundances of the light elements as well as degeneracy among the
independent parameters.  In particular, without a tight independent
constraint on the baryon density, variation in the predicted abundance of
light elements could be largely compensated, at least within observational
uncertainties, by varying $\eta$.

Recently, WMAP has provided an independent measure of $\eta$ \cite{wmap1}
allowing one to use the baryon density as an independent constraint on BBN
instead of inferring its value using the apparent agreement between light
element abundance predictions and observations.  In this vein, prior to the
WMAP results other CMB experiments had sufficiently constrained $\eta$ to
allow tests for the consistency of BBN \cite{Cyburt}.

At the same time recent measurements of the primordial deuterium abundance
($D/H$) using quasar absorption by intervening high redshift systems has
dramatically altered its significance in comparing BBN predictions with
observations \cite{deut2}.

Combining these two new observations gives one a powerful new handle to use
in order to constrain cosmological parameters using BBN predictions.  Here
we explore the impact of these developments on one's ability to constrain
the variability of $G$.  Variation of the gravitational constant was
originally postulated by Dirac \cite{Dirac} and remains a key component of
many theories that seek to resolve various hierarchy problems in particle
physics.  Varying the gravitational constant has significant impact during
BBN.

While previous analyses (for example~\cite{Steigman,Krauss1}) attempted to
utilize all BBN abundance predictions as a way to limit the effect of
uncertainties in $\eta$, as can be seen from the results of \cite{Krauss1},
there is good reason to believe that utilization of deuterium alone, now
that its primordial value has been tightly constrained, might be sufficient
to yield a stronger constraint.  This is because deuterium production is
particularly sensitive to the value of the expansion rate during BBN, which
in turn depends upon the value of $G$. While helium is also sensitive to
this rate, existing systematic uncertainties in its primordial value
confuse the situation.  In addition, while earlier analyses were performed
before it was empirically demonstrated that 3 light neutrino species exist,
this residual uncertainty has now been removed.

It has been known for almost three decades that deuterium is only produced
in significant quantities in the big bang \cite{Reeves,Epstein}.
This, coupled with the fact that deuterium depends sensitively on $\eta$,
makes it an excellent probe of the Universe at the time of BBN (for example see
\cite{Krauss2}).  The detection of deuterium in a number of high redshift
quasar absorption systems allows for a precise determination of the
primordial deuterium abundance. \textcite{deut1} give as the best current
estimate of the primordial deuterium abundance $D/H =
2.78^{+0.44}_{-0.38}\times10^{-5}$ based on the observation of five quasar
absorption systems.  This corresponds to $\log{(D/H)}=-4.556\pm0.064$ where
the errors are assumed to be Gaussian.

The new CMB constraints on $\eta$ are derived from the temperature of the
CMB and a fit of the power spectrum to a set of parameters defining a
cosmological model.  Here we are interested in $\Omega_b h^2$ where
$h\equiv H_0/100\;{\rm km\,s^{-1}\,Mpc^{-1}}$ and $\Omega_b =
\rho_b/\rho_c$, the ratio of the baryon density to the critical density.  A
combined analysis including WMAP, CBI, ACBAR, 2dFGRS, and Lyman $\alpha$
data gives $\Omega_b h^2=0.0224 \pm 0.0009$ for $\Lambda$CDM models for the
two choices of either a power law spectral index or a running spectral
index \cite{wmap1}.  Combined with the photon density,
$n_\gamma=410.4\pm0.9\;\textrm{cm}^{-3}$, as determined from the COBE
temperature measurement \cite{COBE} gives $\eta=\left( 6.13\pm0.25
\right) \times 10^{-10}$ where we again assume Gaussian errors.

\begin{figure}
  \centerline{\includegraphics[width=3.3in]{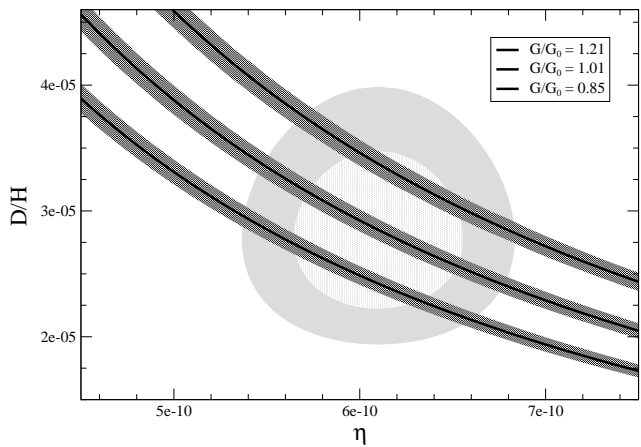}}
  \caption{\label{fig:glimits} Observational limits and theoretical
    expections for $D/H$ versus $\eta$.  The one (light shading) and 2
    (dark shading) sigma observational uncertainties for $D/H$ and $\eta$ are
    shown.  They do not appear as ellipses due to the linear scale in $D/H$
    but logarithmic uncertainties from the observations.  The BBN
    predictions are shown as the solid curves where the width is the $3\%$
    theoretical uncertainties.  Three different values of
    $G_{BBN}/G_0$ are shown.}
\end{figure}

To determine the deuterium abundance as a function of $\eta$ and $G$ from
BBN we use the standard Kawano code \cite{Kawano}.  We use updated reaction
rates \cite{NACRE} as well as used the latest value for the neutron
half-life \cite{pdg}.  Krauss and Romanelli \cite{Krauss3} first
demonstrated the need to incorporate reaction rate uncertainties using
Monte Carlo techniques if one is to properly derive BBN constraints in
general.  The residual uncertainty in the predicted deuterium abundance is
quite small.  We have included this theoretical uncertainty in in the
likelihood analysis as an independent Gaussian with a constant $3\%$
uncertainty at the one sigma level~\cite{BNT,CFO}.  That is,
$\sigma_{\rm BBN} = 0.03 (D/H)$ where
$(D/H) (\eta,G/G_0)$ comes from the Kawano code as noted.
The uncertainties in $G$ today are sufficiently small~\cite{Scherrer} and
have not been included.  In addition, due to the short time interval
associated with BBN we have assumed that $G$ remained constant throughout
the period of light element production.  We have verified that our results
are unchanged if we allow $G$ to vary during this time according to
$G\propto t^{-\alpha}$.  Figure~\ref{fig:glimits} shows the $1$ and $2$
sigma ellipses for the observed values of $D/H$ and $\eta$.  Also shown are
curves for the BBN predictions for $(D/H) (\eta,G/G_0)$ for three
values of $G/G_0$ with the $3\%$ theoretical uncertainties from Monte
Carlos of the reaction rates.

With the assumption that the observational and theoretical errors follow
Gaussian distributions as discussed above, we assign a joint likelihood
function ${\cal L} (D/H,\eta, G/G_0) $ where $D/H$ is a function of $G$ and
$\eta$.  The likelihood distribution for $G$ (which can be normalized to
give the probability distribution) is found by marginalizing over $D/H$ and
$\eta$.  Applying this procedure we find $G/G_0=1.01^{+0.20}_{-0.16}$ at the
$68.3$\% confidence level, $G/G_0=1.01^{+0.42}_{-0.30}$ at the $95$\%
confidence level.

Assuming a monotonic power law time dependence $G\propto t^{-\alpha}$ one
finds the power law index $-0.004 < \alpha <0.005$ at the $68.3$\%
confidence level, $-0.009 < \alpha <0.010$ at the $95$\% confidence level.
In this case one infers $-3\times10^{-13}\;\textrm{yr}^{-1}\left(\dot
  G/G\right)_{\textrm{today}} < 4\times 10^{-13}\;\textrm{yr}^{-1}$. This
is over an order of magnitude stronger than direct constraints that can be
obtained on the variation of $G$ today. 

Our new constraint is about a factor of two stronger than the previous BBN
constraint on $\dot G/G$ \cite{Krauss1}, which is significant, given that
we utilize only deuterium in our analysis, and do not exploit other light
element abundance predictions.  In particular, as observational constraints
on these elements continue to improve, along with further improvements on
the deuterium abundance measurements and the baryon to photon limit one can
expect BBN constraints on variability of $G$ to continue to improve
significantly.

LMK thanks G. Steigman for useful discussions, and pointing out early work
on this subject.

\bibliography{bbn}

\end{document}